\documentclass[12pt]{iopart}

\usepackage{graphics}
\usepackage{epsfig}
\usepackage{bm}

\def\jpb{{\em J.\ Phys.\ B: At.\ Mol.\ Opt.\ Phys.}\ }
\def\pra{{\em Phys.\ Rev.}\ A }
\def\pre{{\em Phys.\ Rev.}\ E }
\def\prl{{\em Phys.\ Rev.\ Lett.}\ }

\def\beq{\begin{equation}}
\def\eeq{\end{equation}}

\def\reff#1{(\ref{#1})}

\def\subsc#1{{\mbox{\rm\scriptsize #1}}}

\def\Wcmcm{\mbox{\rm Wcm$^{-2}$}}

\def\Nouter{N_\subsc{outer}}
\def\Zavcs{Z_\subsc{avcs}}
\def\Fignition{F_\subsc{ignition}}
\def\Xcm{X_\subsc{cm}}
\def\tth{t_\subsc{th}}

\def\N3d{N_\subsc{3D}}

\def\rw{r_\subsc{WS}}

\def\vekt#1{\bm{#1}}

\def\vektr{\vekt{r}}

\def\vektR{\vekt{R}}

\def\vekte{\vekt{e}}
\def\vektE{\vekt{E}}

\def\vektp{\vekt{p}}
\def\vektP{\vekt{P}}

\def\vektJ{\vekt{J}}

\def\rtilde{\tilde{r}}

\def\ptilde{\tilde{p}}

\def\hamop{{\cal{H}}}

\def\Edach{\hat{E}}

\def\Ehat{\Edach}

\def\Zeff{Z_{\mbox{\rm\scriptsize eff}}}
\def\Rcluster{R_{\mbox{\rm\scriptsize cluster}}}

\def\energy{{\cal{E}}}

\def\calV{{{\cal V}}}

\begin{document}

\title[Small rare gas clusters in laser fields]{Small rare gas clusters in laser fields: ionisation and absorption at long and short laser wavelengths}

\author{D.\ Bauer}

\address{Max-Born-Institut, Max-Born-Strasse 2a, 12489 Berlin, Germany

E-mail: bauer@mbi-berlin.de}

\begin{abstract}
The interaction of short laser pulses with small rare gas clusters is investigated by using a microscopic, semi-classical model with an explicit treatment of the inner-atomic dynamics. Field and collisional ionisation as well as recombination are incorporated self-consistently so that the use of rates for these processes could be avoided. The laser absorption and ionisation mechanisms in clusters at near-infrared ($800$\,nm) and VUV wavelength ($100$\,nm) are analysed.

\bigskip
\noindent (Figures in this article are in colour only in the electronic version)

\bigskip

\end{abstract}


\submitto{\JPB}

\maketitle

\section{Introduction}
Clusters in intense laser fields were found to absorb laser energy particularly well. The ions inside such clusters typically assume higher charge states than expected from a single, isolated atom in the same laser pulse. The reason lies in the complex interplay between the electric field inside the cluster and ionisation, both ``inner ionisation,'' that is, the removal of electrons with respect to individual ions, and ``outer ionisation'' (i.e., removal of electrons from the cluster as a whole).
In experiments at wavelengths $\geq 248$\,nm highly energetic electrons \cite{shao,springelecs,kuma}, ions \cite{ditmirenature,ditmirePRA,lezius,leziusII,springate}, photons \cite{mcpherson,dobosz,ditmireJPB,teravet}, and neutrons originating from nuclear fusion \cite{zweibackDD} were observed (see Refs.\ \cite{review,reviewII} for reviews). The first Xe cluster experiments in the VUV regime at high laser intensities $>10^{12}$\,\Wcmcm\ were performed at the DESY free-electron laser (FEL), Hamburg, Germany \cite{wabnitz}, showing higher charge states than expected from single atoms as well, despite the tiny quiver energies of electrons at such short wavelength.

The interaction of moderately sized clusters with strong laser fields is commonly simulated within the framework of molecular dynamics where the inner ionisation is incorporated via rate equations or a sequential, single active electron-approach (see, e.g., \cite{rosepetru,lastI,lastII,ishika,saalmann,siedschlagI,siedschlagII,siedschlagIII}). Instead, in this work a semi-classical approach with an explicit treatment of the multi-electron inner-atomic dynamics is pursued.  In that way field or collisionally induced ionisation, recombination, non-sequential ionisation, and other inelastic processes are included self-consistently.

The paper is organised as follows. In section \ref{scm} the semi-classical cluster model is introduced before in section \ref{rd} the numerical results, especially the ionisation mechanism and the energy absorption in clusters at $800$ and $100$\,nm are discussed. We conclude in section \ref{concl}.

\section{Semi-classical model} \label{scm}
The rare gas cluster consisting of $N_a$ ions of mass $M$ at the positions $\vektR_i$, $1\leq i \leq N_a$,  and $Z$ ``active'' electrons per ion at the positions $\vektr_j$, $1\leq j \leq ZN_a$ in a laser field $\vektE(t)$ is modelled by the Hamiltonian
\begin{eqnarray} \hamop(\vektR,\vektP;\vektr,\vektp;t)&=&\sum_{i=1}^{N_a} \frac{\vektP_i^2}{2M} +  \sum_{j=1}^{Z N_a} \frac{\vektp_j^2}{2} + \vektE(t)\cdot\Bigl(\sum_{j=1}^{Z N_a} \vektr_j - Z \sum_{i=1}^{N_a} \vektR_i \Bigr) \label{hamiltonian}\\
&&  + \sum_{i=1}^{N_a}\sum_{j=1}^{Z N_a} \Bigl(  V_H(\rtilde_{ij},\ptilde_{ij}) - \frac{Z}{\vert\vektR_i-\vektr_j\vert} \Bigr)    \nonumber \\
&& + \sum_{i=1}^{N_a} \sum_{k=1}^{i-1} \frac{Z^2}{\vert\vektR_i-\vektR_k\vert} + \sum_{j=1}^{ZN_a} \sum_{l=1}^{j-1} \frac{1}{\vert\vektr_j-\vektr_l\vert} \nonumber 
\end{eqnarray} 
(atomic units are used until noted otherwise).
Since ``classical atoms'' with more than one electron are generally unstable, a momentum-dependent potential
\beq
\calV(r,p,\xi,\alpha,\mu)=\frac{\xi^2}{4\alpha r^2 \mu} \exp\left\{ \alpha \left[ 1- \left(\frac{rp}{\xi}\right)^{\!\!4} \right]\right\}
\label{momdeppot} \eeq
is introduced \cite{kw} that approximately enforces the Heisenberg uncertainty relation when applied in the form
$ 
V_H(\rtilde_{ij},\ptilde_{ij})=\calV(\rtilde_{ij},\ptilde_{ij},\xi_H,\alpha_H,\mu_{ei}) 
$, where 
$ \rtilde_{ij}= \vert \vektr_i-\vektR_j \vert$, $\ptilde_{ij}=\left|{(M \vektp_i- \vektP_j)}/{(1+M)}\right|$ are relative position and momentum, respectively, and $\mu_{ei}$ is the reduced mass of electron and ion.
Spin effects are neglected but it is worth noticing that Pauli-blocking could be incorporated in the model \cite{cohen,beckwilets,cohen2000}. The bigger the ``hardness parameter'' $\alpha_H$ is chosen, the more severely 
the uncertainty relation $\rtilde_{ij}\ptilde_{ij}\geq\xi_H$ is fulfilled. However, big $\alpha_H$ render the equations of motion stiff so that in practical calculations a compromise has to be made.
$\xi_H$ is used as a free parameter for adjusting the relevant physical properties (e.g., ionisation potentials) of the atomic species under study. 

The results presented in this work are obtained by solving the equations of motion for all electrons and ions corresponding to the Hamiltonian \reff{hamiltonian} for both a model Xe atom and a model Xe$_{54}$ cluster. Only the $Z=6$ electrons in the 5p valence shell of each Xe atom are considered ``active.'' This is a reasonable assumption for the laser intensity regime discussed in the present work and significantly reduces the numerical effort. $\alpha_H=2$ and $\xi_H=2.33$ was chosen which leads to the six ionisation potentials $\energy_I=0.16$, $0.55$, $1.11$, $1.63$, $2.22$, $2.65$ for an isolated Xe atom in this model while the real ionisation potentials are (see, e.g., \cite{cowley,cowan}) $\energy_I=0.45$, $0.77$, $1.18$, $1.69$, $2.09$, $2.64$. The first two ionisation potentials of the model are in poor agreement with the correct values. Unfortunately, we found it not possible to tune all six ionisation potentials to the correct values by varying the only free parameter $\xi_H$ (the parameter $\alpha_H$ hardly affects these values \cite{beckwilets}). Since it is the generation of the higher charge states in clusters that is of particular interest to us, $\xi_H$ was chosen to optimise the higher ionisation potentials.

\section{Results and discussion} \label{rd}
\subsection{Single-atom results}
\begin{figure}
\mbox{\qquad\qquad\qquad}\includegraphics[scale=0.7]{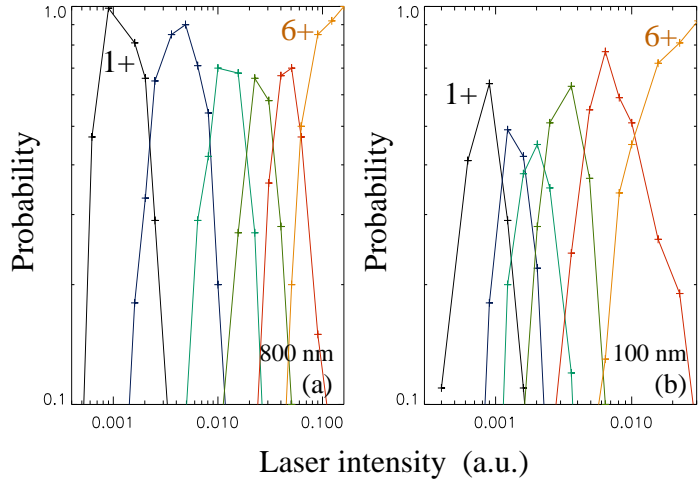}
\caption{\label{isolatedatomresult}  Probabilities to obtain charge states 1+, 2+, ..., 6+  vs laser peak intensity for the Xe model atom in a trapezoidally shaped laser pulse of $T\approx 42$\,fs duration and wavelength $800$\,nm (a) and $100$\,nm (b). Multiply the laser intensity by $3.51\times 10^{16}$ to obtain the corresponding value in \Wcmcm.  }
\end{figure}

The stationary ($\dot{\vektr}=\dot{\vektp}=0$) electronic configuration of the single model atom is face-centred cubic, each of the six electrons having the same distance $\vert\vektr\vert=1.56$ to the ion located in the centre of the cube. The canonical momentum is finite, $\vert\vektp\vert=1.49$.
In order to investigate the enhanced (or reduced) ionisation in clusters one needs to know the probabilities for the various charge states of the single, isolated atom. In Fig.\,\ref{isolatedatomresult} the probabilities to find a certain charge state  are presented for the two laser wavelengths $800$ and $100$\,nm. The laser pulse was of trapezoidal shape with an $\approx 8$\,fs up and down-ramp (corresponding to $3$ and $24$ laser cycles at $800$ and $100$\,nm, respectively) and a $T_\subsc{const}=26.5$\,fs ($10$ and $80$ cycles, respectively) flat top. 
The charge state distributions in Fig.\,\ref{isolatedatomresult} were obtained by simulating an ensemble of $100$ randomly rotated single atoms in the laser field. Note that the electron dynamics in the laser field and thus the final charge state in general depends on the orientation of the initial, classical electronic configuration with respect to the laser polarisation $\vekte_x$.   

The so-called ``saturation intensity'' for a certain charge state is defined as the laser intensity where the charge state reaches its maximum probability before it decreases in favour of the next higher charge state. The saturation intensities for charge states $Z$ higher than 2+ in Fig.\,\ref{isolatedatomresult}a agree well with what is expected from the Bethe rule \cite{bethe} $I\subsc{sat}=\energy_I^4/(16 Z^2)$ while the first two charge states appear at higher intensities. 
This is likely due to the incomplete screening of the full nuclear charge by the other electrons and a non-negligible down-shift of the binding energy (classical Stark effect) for the first two electrons in our model. Note that in the derivation of the Bethe rule a Stark shift is not taken into account. However, in real experiments on Xe in $800$\,nm laser light \cite{laro} also the saturation intensities for the lowest charge states obey the Bethe rule. This means that, by fortunate coincidence, the too low binding energies of the first two electrons are partly compensated by incomplete screening and classical Stark effect so that the first two saturation intensities in our model are less underestimated as one could expect from the ionisation potentials alone. 
At $100$\,nm (Fig.\,\ref{isolatedatomresult}b), the maximum probabilities to find successive charge states in our model Xe atom lie closer together so that broader charge state distributions arise than at $800$\,nm. 

\begin{figure}
\mbox{\qquad\qquad\qquad}\includegraphics[scale=0.5]{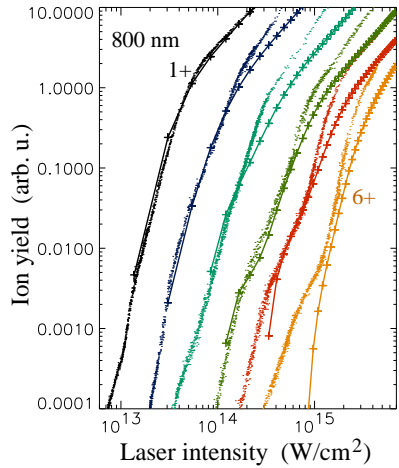}
\caption{\label{yieldcomp}  Numerically determined and focus averaged ion yields vs laser intensity (+). Dots are experimental data from \cite{laro}. The experimental intensities were multiplied by a factor of $0.4$\,. }
\end{figure}

In Fig.\,\ref{yieldcomp} focus-averaged charge state distributions at $800$\,nm are compared with the experimental data in \cite{laro} for $200$\,fs laser pulses. The overall agreement is surprisingly good. Note, however, that the experimental laser intensity had to be multiplied by a factor $0.4$ for matching our model results so that only good {\em qualitative} agreement can be claimed. 

At $100$\,nm the energy of a photon is close to the first ionisation potential of real Xe. It would be not surprising if our semi-classical model breaks down at such short wavelengths where ionisation becomes a single-photon process. Although the model  possesses eigenfrequencies, these have in general nothing to do with quantum mechanical excitation or ionisation energies. 
First results for $98$\,nm were obtained at the DESY-FEL in Hamburg, Germany  \cite{wabnitzXeatoms} (note that in our semi-classical approach it does not matter whether one uses $98$ or $100$\,nm wavelength). A comparison  suggests that our model overestimates the saturation intensities. However, the uncertainty in the peak laser intensity and the spiky structure of the FEL pulse inhibits a straightforward comparison between theory and experiment. Moreover, it should be emphasised that not the absolute values for threshold or saturation laser intensities are the main focus of our studies but the {\em relative} differences between single atom and cluster ionisation.

\subsection{Cluster results}
Clusters consisting of $N_a=54$ model Xe atoms in the laser pulse of shape and duration as described in the previous subsection were simulated. The initial cluster configuration was assembled by attaching randomly rotated single-atom electron configurations to the ions. The ion positions $\vektR_i$ were taken from normalised Lennard-Jones cluster calculations \cite{wales,waleswww} with each radial vector multiplied by $7.127$ so that a Wigner-Seitz radius $\rw\approx 4$ was obtained.  
More details on the actual implementation of our cluster model can be found in Ref.~\cite{bauerappl}.

\begin{figure}
\mbox{\qquad\qquad\qquad}\includegraphics[scale=0.7]{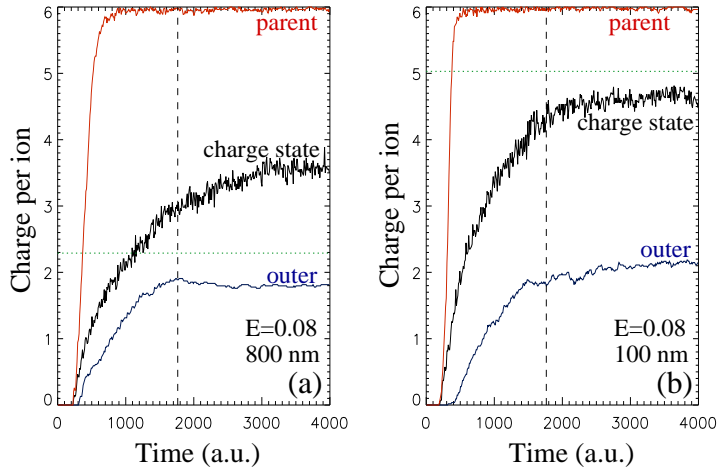}
\caption{\label{ioni}  Cluster ionisation vs time for the peak laser fieldstrengths $\Ehat=0.08$ and  $\lambda=800$ (a) and $100$\,nm (b). The average number of electrons per ion that left their parent atom (``parent''), any atom (``charge state''), or the entire cluster (``outer'') are presented. Vertical dashed lines indicate the end of the laser pulse. Horizontal green-dotted lines mark the charge state of an isolated atom after the same laser pulse. See text for discussion.}
\end{figure}

\subsubsection{Ionisation mechanism}
In order to get a deeper insight into the cluster ionisation dynamics it has proven useful to distinguish between three groups of electrons. First, there are those electrons that left their parent atom.
We consider an electron to contribute to this ``parent atom ionisation'' if it is more than a Wigner-Seitz radius away from its parent atom. 
Electrons in the second group are more than a Wigner-Seitz radius away from {\em any} ion and thus define the averaged ion charge state
\beq \Zavcs=\frac{1}{N_a}\Bigl( ZN_a-\sum_{i=1}^{N_a} \sum_{j=1}^{ZN_a} \Theta (\rw-\vert\vektr_j-\vektR_i\vert) \Bigr)  \label{avcs}\eeq
where $\Theta$ is the step-function.

Finally, there are those electrons that are far away from the entire cluster. We count these electrons by calculating 
\beq \Nouter=\sum_{j=1}^{ZN_a}\Theta(r_j-\max\{ R_i\}-\rw)\label{outer}\eeq
which equals the total charge of the cluster that is left behind.
The charged cluster is unstable and Coulomb-explodes. Under the conditions investigated in this work, a typical time within which the cluster doubles its radius is several tens of femtoseconds. In the definition \reff{outer} the cluster radius is taken as $\Rcluster=\max\{ R_i\}+\rw$ where $\max\{ R_i\}$ is the distance from the outermost ion to the cluster centre.  

In Fig.~\ref{ioni} the parent atom ionisation, the average charge state, and outer ionisation are plotted vs time for the laser intensity $I=\Ehat^2=6.4\times 10^{-3}$, corresponding to $2.2\times 10^{14}$\,\Wcmcm, and the two wavelengths $\lambda=800$ and $100$\,nm. Common features in both results are the fast removal of {\em all} active electrons from their parent atom and the moderate outer ionisation ($\approx 2$ electrons per ion are removed from the entire cluster). 
The removal of all six electrons from their parent atom might be surprising in view of the fact that single, isolated atoms in the same field loose not even three electrons (see horizontal, green-dotted line in Fig.~\ref{ioni}a). At lower laser intensities this discrepancy between single atom ionisation and parent atom ionisation in clusters is even more pronounced. In fact, we found that it is in general sufficient that the single, isolated atom looses one electron to trigger complete parent atom ionisation in the cluster. However, parent atom ionisation is {\em not} an observable! Electrons are indistinguishable, and, indeed, {\em other} electrons are trapped by the ion. The average charge states---depending on the balance between ionisation and recombination---are significantly lower in Fig.~\ref{ioni}. Looking at the particle trajectories one infers that the electrons hop from one ion to the other, undergoing inelastic collisions and getting temporarily trapped. This is expected because a low temperature, strongly coupled plasma of electron density $n_e$ and Debye-length $\lambda_D$ is formed whose  plasma parameter $1/(n_e\lambda_D^3)$ is close to unity  or even higher. Consequently, a Debye-sphere is populated by only a few electrons, and the potential energy of the electrons is comparable to their kinetic energy.

One may object that complete parent atom ionisation suggests that more active electrons per atom should be incorporated in the simulation in the first place. However, what is important is that the charge states
\beq \Zeff^{(i)} = Z - \sum_{j=1}^{ZN_a} \Theta (\rw-\vert\vektr_j-\vektR_i\vert) \label{cs}\eeq
remain smaller than the highest allowed charge state $Z$. Whether an electron screens (or is bound to) its parent atom or another one is insignificant. Hence, increasing the number of active electrons per atom while $\Zeff^{(i)}<Z$ introduces  an unnecessary large overhead of bound electrons to be propagated.   
The stronger coupled the plasma is, the greater the difference between the ``parent atom ionisation degree'' and the charge states \reff{cs} is expected.

The average charge state $\Zavcs$ (indicated by ``charge state'' in Fig.~\ref{ioni}) was calculated geometrically by counting all electrons farther away than $\rw=4$ from any ion.
This geometric concept is meaningful only when the cluster has sufficiently expanded so that it becomes unlikely that free electrons occupy accidentally the test spheres of radius $\rw$ around the ions. In fact, the cluster expansion is the reason why the charge states appear to increase even after the end of the laser pulse (indicated by the dashed vertical lines). Hence, only the asymptotic values of the ``charge state''-curves in Fig.~\ref{ioni} are the expected average charge states that would be measured in a real experiment.
In other words: when the ions are well separated (as it is the case in an experiment, shortly before they hit the detector) all electrons within a sphere of radius $\rw$ around the ions can be safely considered bound so that \reff{cs} is indeed the charge state that would be measured, and, consequently,  $\Zavcs=\sum_{i=1}^{N_a} \Zeff^{(i)}/N_a$ would be the average charge state provided that all ion species are detected with the same efficiency. 
In order to figure out whether the cluster ionises more efficiently than the single, isolated atom, the average charge state $\Zavcs$ can be compared with the corresponding  results presented in the previous subsection.
These single atom results are included in Fig.~\ref{ioni} (horizontal, green-dotted lines).

\begin{figure}
\mbox{\qquad\qquad\qquad}\includegraphics[scale=0.6]{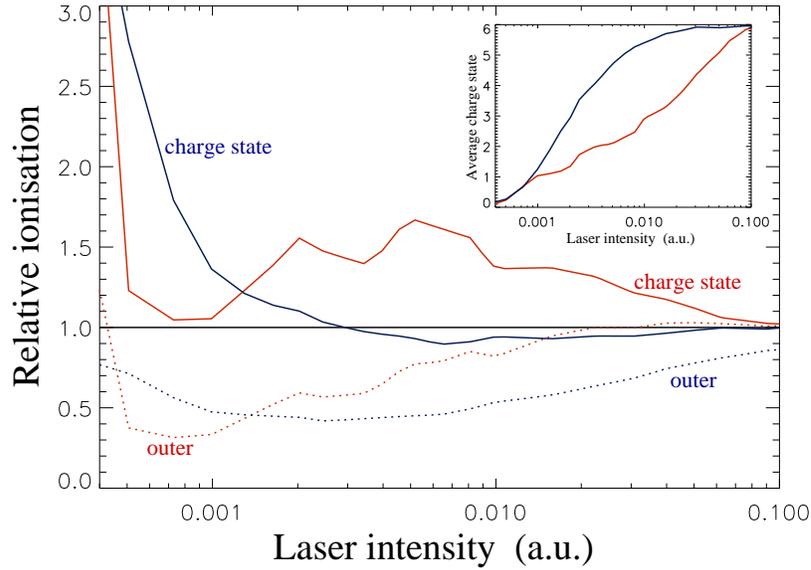}
\caption{\label{relioni}  Cluster ionisation normalised to the single, isolated atom ionisation (see inset) vs laser intensity for the two wavelengths $\lambda=800$ (red) and $100$\,nm (blue). Average charge states (solid lines) and outer ionisation per cluster atom (dashed) are shown.  See text for discussion.  }
\end{figure}

It is seen in Fig.~\ref{ioni} that, for the particular laser intensity chosen,  at $100$\,nm the cluster ionisation is less efficient than the single-atom ionisation. The ``relative ionisation,'' that is, the average charge state in the cluster divided by the corresponding single atom charge state, is shown in  Fig.~\ref{relioni} as a function of the laser intensity. A value above unity indicates that ionisation in the cluster is more efficient than ionisation of the isolated atom.
At both laser wavelengths the relative ionisation in the cluster is particularly high when the first free electrons are generated. The ionisation rate for the very early electrons in the cluster necessarily equals the one for isolated atoms since there is not yet the potential from the other ions. However, as soon as these very early electrons leave, the cluster charges up, leading to strong electric fields, especially at  the cluster boundary, that enhance further inner ionisation (``ionisation ignition'' \cite{rosepetru}, details follow below). This is the reason for the strong relative ionisation at intensities $I< 6\times 10^{-4}$ for both wavelengths.

At higher laser intensities inner ionisation increases but the electrons are less efficiently transported away from the cluster as a whole---especially at short wavelengths because the quiver amplitude of the freed electrons is low then so that outer ionisation can proceed through thermionic emission only. 
The electron temperature that is reached is sensitive to the pulse duration. For the $42$\,fs-pulses and the laser intensities in the present work the electron temperatures at $100$\,nm always remain below $1.7$ atomic units. During the cluster expansion, the electrons trapped by the cluster potential are cooled, and recombination leads to only moderate final charge states (the higher, transient charge states should be experimentally accessible by measuring the emitted X-rays).
As a result, the relative ionisation at $100$\,nm drops below unity, that is, the atoms inside the cluster ionise---on average---less efficiently than an isolated atom does. 
At $800$\,nm and the same laser intensity relative ionisation remains above unity, mainly because single atom ionisation is lower (see inset in Fig.~\ref{relioni}) and the quiver energy suffices to free the electrons from the cluster potential (see red, dotted outer ionisation-curve in Fig.~\ref{relioni}).
Finally, at sufficiently high laser intensities all active electrons are removed from both the cluster atoms and the isolated atom so that relative ionisation approaches unity.

\begin{figure}
\mbox{\qquad\qquad\qquad}\includegraphics[scale=0.92]{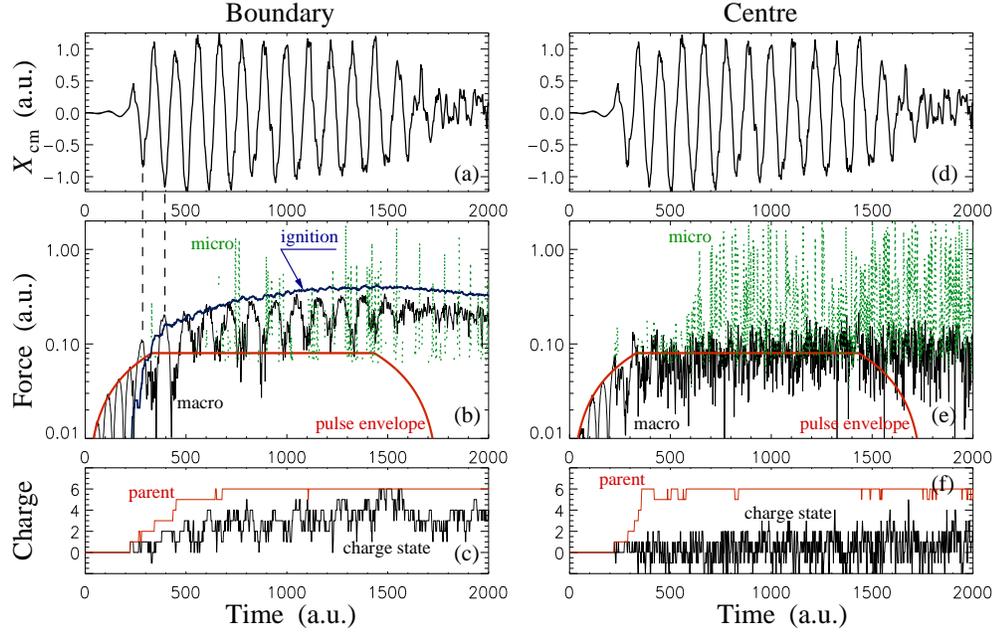}
\caption{\label{fieldsn1}  Panels (b) and (e) illustrate the absolute values of the external forces vs time, as seen by an electron bound to an ion at the cluster boundary and in the cluster centre, respectively. The electric field envelope of the laser pulse ($\Ehat=0.08$, $\lambda=800$\,nm, marked ``pulse envelope''), the force exerted by all electrons and ions outside the test sphere of radius $\rw=4$ plus the force of the laser (``macro''), and the force exerted by all electrons passing the test sphere within a time $\tth=8$ (``micro'') are shown. In (b) the ``ignition'' force expected due to the positively charged cluster is included as well.   The centre of mass of all electrons inside the cluster $\Xcm$ is plotted in (a) and (d). Parent atom ionisation and charge state of the cluster atom under consideration are shown in panels (c) and (f). }
\end{figure}

In order to analyse the origin of the efficient ionisation inside the cluster in more detail it is useful to look at the external forces a bound electron experiences. This is done in Fig.~\ref{fieldsn1} for the two cases of an electron bound to an ion at the cluster boundary and to an ion sitting in the cluster centre. The laser parameters were as in Fig.~\ref{ioni}a. The forces experienced by the electron are divided into a macro field-part which is made up by the laser and all electrons and ions that are outside a test sphere of radius $\rw=4$ around the ion of interest, and into a micro field-part originating from electrons passing through the test sphere within $\tth=8$ atomic time units. The threshold time $\tth$ is introduced in order to estimate the importance of collisional ionisation of the type
$ e + A^{z+} \longrightarrow 2e + A^{(z+1)+}$ without allowing processes where the incoming electron is captured to affect the micro field (like $e + A^{z+} \longrightarrow e' + A^{z+}$, for instance). Other {\em ions} never pass through the test sphere because the ions do not overtake each other (see below for an explanation).

From Fig.~\ref{relioni} one infers that for the laser parameters used in Fig.~\ref{fieldsn1}, $I=\Edach^2=6.4\times 10^{-3}$ and $\lambda=800$\,nm, the average cluster ionisation is more efficient than in isolated atoms. Fig.~\ref{fieldsn1}b explains why. During the first four laser cycles the test atom exclusively ``sees'' the laser field because ionisation is still very low so that there are no free electrons around. Then the first electrons are removed (see  the parent atom ionisation in Fig.~\ref{fieldsn1}c) and the macro field significantly exceeds the value one would expect from the laser field alone. Looking at the centre of mass in laser field direction (Fig.~\ref{fieldsn1}a)  of all electrons inside the cluster, it is seen that the pronounced maxima in the macro field occur at times where most electrons are on the opposite side of the cluster so that the screening of the positive cluster charge is low  (see dashed lines to Fig.~\ref{fieldsn1}a). Half a laser cycle later, when most electrons move to the side of the cluster where the test atom is located, the macro field drops dramatically. This clearly shows that a {\em dynamical} version of ionisation ignition should be adopted, as proposed in Ref.~\cite{baumacch}. As outer ionisation increases, the maxima of the macro field increase as well. This is due to the positive ion background that pulls electrons from the cluster boundary into the interior, i.e., the ``standard'' ionisation ignition mechanism. The force
$ \Fignition={\Nouter}/{R^2}$,
where $R$ is the radial position of the test atom under consideration and $\Nouter$ is the cluster charge as defined in \reff{outer},
is included in plot (b). It is seen that the overall temporal evolution of the macro field is in good agreement with $\Fignition$. Deviations, especially at late times, are due to the inappropriate assumption of a homogeneously charged sphere for an expanding cluster consisting of only $54$ ions. The micro field, caused by electrons passing through the test sphere, does not contribute to a net increase of the charge state but only to fluctuations (see Fig.~\ref{fieldsn1}c).

The test atom in the centre clearly experiences no ionisation ignition. Although this atom looses all its ``initial'' electrons (see ``parent''-curve in Fig.~\ref{fieldsn1}f) the overall charge state remains low, in fact, lower than for an isolated atom in the same laser field.  This shows again that calculating charge states from parent atom ionisation would lead to wrong results. The micro field is stronger in the cluster centre because the electron density is higher there. However, the micro field fluctuations again do not increase the net charge state on longer time scales. Hence, collisional ionisation is not responsible for the increased ionisation at $800$\,nm in Fig.~\ref{relioni}, neither for atoms at the cluster boundary nor for atoms in the centre.

\begin{figure}
\mbox{\qquad\qquad\qquad}\includegraphics[scale=0.92]{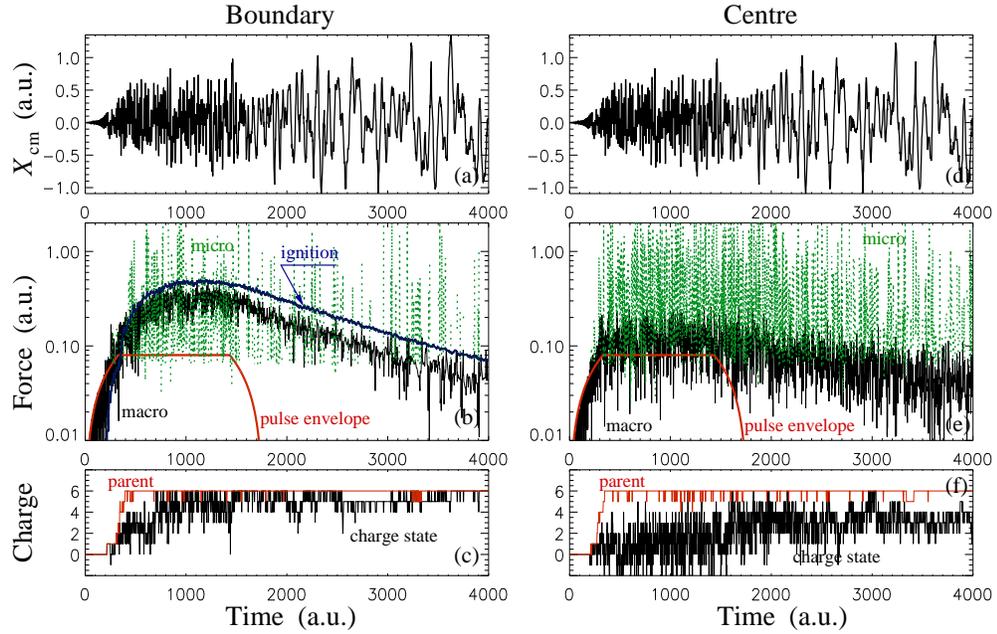}
\caption{\label{fieldsn8}  Same as in Fig.~\ref{fieldsn1} but for $\lambda=100$\,nm. Note that the time interval shown is twice as long now.  }
\end{figure}

From Fig.~\ref{relioni} one infers that at the shorter wave length $\lambda=100$\,nm and $I=6.4\times 10^{-3}$ the average charge state inside the cluster is slightly less than for the isolated atom. In Fig.~\ref{fieldsn8} the external forces experienced by a bound electron at the cluster boundary and in the centre are presented for this case. Because of the higher laser frequency the centre of mass-excursion is less pronounced so that collective electron dynamics inside the cluster does not build up. Note that the centre of mass-excursion is actually less when the laser is on than it is after the laser pulse, as it is expected for a driven harmonic oscillator when the driving frequency is higher than the eigenfrequency (i.e., the Mie plasma frequency in our case).  Outer ionisation during the laser pulse proceeds through collisional heating (see section~\ref{energabsorb} below) so that the cluster charges up and the macro field again nicely follows $\Fignition$ in Fig.~\ref{fieldsn8}b. The final charge state is maximal, i.e., $6$, but one has to keep in mind that also the isolated atom looses already $\approx 5$ electrons in this laser pulse. The number of electrons inside the test sphere around the ion in the cluster centre strongly fluctuates, slightly drops (corresponding to an increase of $\Zeff$ in Fig.~\ref{fieldsn8}f) after the laser pulse because of the previously mentioned localisation effect of the laser and cluster expansion,  but remains below the isolated atom result.

\begin{figure}
\mbox{\qquad\qquad\qquad}\includegraphics[scale=0.45]{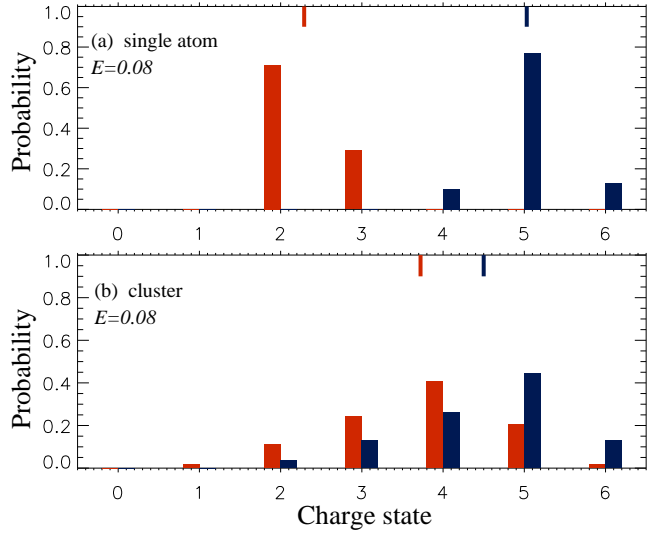}
\caption{\label{csdistrH}  Charge state distributions at $I=\Edach^2=6.4\times 10^{-3}$, $\lambda=800$ (red) and $100$\,nm (blue) for the single, isolated atom (a) and the cluster consisting of $54$ atoms (b). The mean values are indicated by the narrow bars at the top of each panel.    }
\end{figure}

The fact that the charge states vary from high (at the cluster boundary) to low (in the cluster centre) already suggests that the charge state distribution should be much broader than in the single, isolated atom-case. This is confirmed in Fig.~\ref{csdistrH} where the charge state distributions for both laser wavelengths can be compared with each other and the corresponding single, isolated atom-results. At the shorter wavelength the average charge state in the cluster falls behind the isolated atom ionisation because the electron temperatures reached are not high enough to yield sufficient outer ionisation through thermionic emission. As a result, the electron density inside the cluster potential remains high and, through recombination, leads to a finally broad charge state distribution with a mean value lower than in the single atom-case.  

Another consequence of the fact that the charge states increase from the cluster centre towards the cluster boundary is that during the Coulomb explosion the ions do not overtake. In the definition of the micro field above ions therefore needed not be taken into account.

From Fig.~\ref{relioni} it can be inferred that in our model only for $I<0.002$ the average cluster ionisation is more efficient than the isolated atom ionisation. It might be this regime into which  the enhanced ionisation measured in Xe clusters at $98$\,nm for laser intensities up to $7\times 10^{13}$\,\Wcmcm\ \cite{wabnitz} falls into.   It would be interesting to verify experimentally whether at higher laser intensities relative ionisation in small clusters really decreases, as it is suggested by our results in Fig.~\ref{relioni} for $I \approx 0.006$.

\subsubsection{Energy absorption} \label{energabsorb}
The energy absorbed by the cluster in the laser field $\vektE(t)$ is
\beq \Delta\energy = \hamop(T)-\hamop(0) = \sum_{i=1}^{N_a}\Delta\energy_i + \sum_{j=1}^{ZN_a} \Delta\energy_j \label{totabs}\eeq
where $T$ is the pulse duration, and 
\beq \Delta\energy_i = Z \int_0^T \!\!  \vektE(t)\cdot\dot{\vektR}_i\, dt, \quad \Delta\energy_j = - \int_0^T \!\!  \vektE(t)\cdot\dot{\vektr}_j\, dt \eeq
are the absorbed energies of the individual ions and electrons, respectively.
Hence it is necessary for laser energy to be absorbed that the total current $\vektJ=Z \sum_i\dot{\vektR}_i - \sum_j\dot{\vektr}_j$ acquires a phase lag $\neq \pm\pi/2$ with respect to the driving field $\vektE(t)$ so that $\Delta\energy = \int_0^T \vektE(t)\cdot\vektJ\, dt \neq 0$. There are several mechanisms that can lead to dephasing: electron-ion collisions (i.e., inverse bremsstrahlung), electrons ``colliding'' with the cluster boundary \cite{review,megi}, and electrons colliding with the cluster as a whole \cite{kostyukov,smirnkrainpop}.
In the following the relative importance of these mechanisms is analysed for our case of a small cluster in a low- and high-frequency laser field.

\begin{figure}
\mbox{\qquad\qquad\qquad}\includegraphics[scale=0.55]{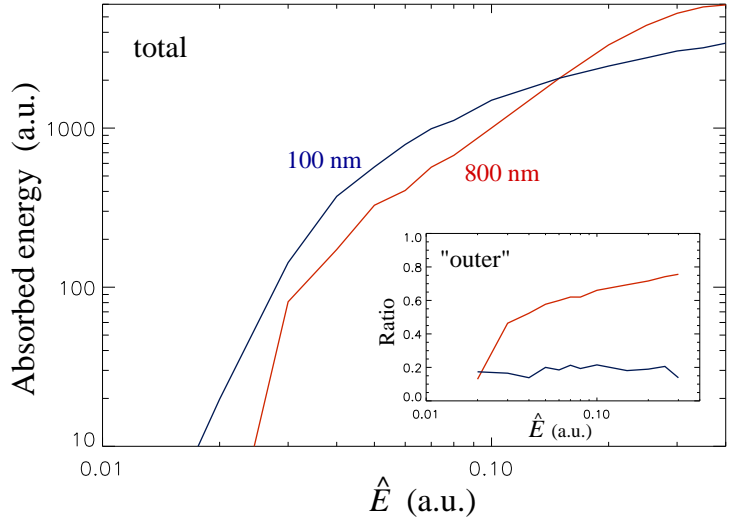}
\caption{\label{absorb}  Laser energy absorbed by the Xe$_{54}$ model cluster for the two laser wavelengths vs the laser field amplitude $\Ehat$.  The fraction $\Delta\energy_\subsc{oa}/\Delta\energy$, 
contributed by electrons while they are farther away than $\rw=4$ from any ion (``outer absorption''), is shown in the inset.    }
\end{figure}

Fig.~\ref{absorb} shows the total absorbed energy \reff{totabs} vs the peak field strength of the laser. Absorption is more efficient for the short wavelength up to $\Ehat\approx 0.13$. This is  because inner ionisation is stronger for the high frequency laser field so that both electron density and average charge states are higher than at $\lambda=800$\,nm. However, the long wavelength-result overtakes around $\Ehat\approx 0.13$. This can be attributed to the increasing importance of electron-cluster and electron-cluster boundary collisions, as it is evidenced in the inset in Fig.~\ref{absorb} where the ratio $\Delta\energy_\subsc{oa} / \Delta\energy$ with
\beq \Delta\energy_\subsc{oa} = - \int_0^T \!\!\!  \vektE(t)\cdot\sum_{j=1}^{ZN_a}\dot{\vektr}_j\prod_{i=1}^{N_a} \Theta(\vert\vektr_j-\vektR_i\vert - \rw)\, dt \label{outerabsorb} \eeq
is plotted. The ``outer absorption'' $\Delta\energy_\subsc{oa}$ measures the laser energy that is absorbed by mechanisms different from standard inverse bremsstrahlung through electron-ion collisions because only electrons while they are farther away than $\rw=4$ from any ion contribute in \reff{outerabsorb}. Since the maximum excursion of free electrons in a laser field is $\sim \Edach/\omega^2$ it is expected that cluster boundary-effects are more important at $800$\,nm. In fact, even at the highest field amplitude shown in the inset of Fig.~\ref{absorb} the excursion $\Edach/\omega^2=1.4$ at $100$\,nm is still much smaller than the initial cluster radius $15.4$ so that it is clearly the thermal velocity that dominates the dynamics at $100$\,nm.

Also connected to the more efficient absorption at $800$\,nm in stronger fields is the fact that freed electrons are rescattered from the entire cluster while they oscillate with large amplitude in the laser field. At $100$\,nm instead the electrons leave the cluster through thermionic emission, and the quiver amplitude in the laser field is too small for driving them back to the cluster. Consequently, the freed electrons  drift away and are thus lost for further energy absorption.

\begin{figure}
\mbox{\qquad\qquad\qquad}\includegraphics[scale=0.6]{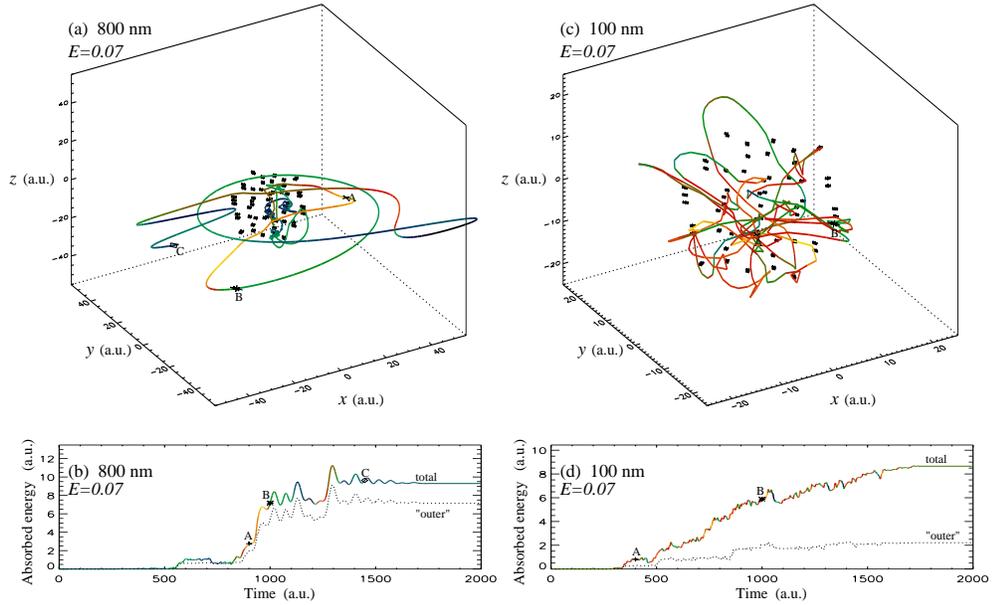}
\caption{\label{trajabsorb}  Trajectory (a,c) and absorbed energy vs time (b,d) of the cluster electron that contributes most to the total energy absorbed by the cluster at $800$ (a,b) and $100$\,nm (c,d) for a laser field amplitude $\Ehat=0.07$. Colour indicates whether the cycle-averaged absorption rate is low (dark colours) or high (light colours). The ``outer absorption'' contribution (dashed) is also included in plots (b) and (d). For the sake of clarity, the electron trajectories in (a) and (c) are shown only for times $t\in[0,1450]$ and $t\in[400,1000]$, respectively. }
\end{figure}

Fig.~\ref{trajabsorb} illustrates with sample trajectories when and where electrons absorb predominantly laser energy. The absorption rate averaged over a laser cycle $T$, 
\beq \dot{\energy}_j(t)=- \frac{1}{T} \int_{t-T/2}^{t+T/2} \!\!  \vektE(t')\cdot\dot{\vektr}_j\, dt', \eeq 
is a measure for the efficiency of laser energy absorption and is used to colour-code the trajectories and curves in Fig.~\ref{trajabsorb}.

In the long wavelength case collisions with the cluster as a whole clearly dominate. The biggest jump in the absorbed energy in Fig.~\ref{trajabsorb}b occurs between points A and B where the electron crosses rapidly the cluster and large-angle electron-ion collisions are obviously absent.    Consequently, the fraction of ``outer absorption'' is particularly high (see dashed curve in  Fig.~\ref{trajabsorb}b). Note, that the time the electron takes to pass the cluster from A to B is of the order of a laser period. Hence, the so-called impact approximation (collision time $\ll$ laser period) does not apply, and efficient energy absorption is possible also in small angle scattering events.

The electron trajectory in the short wavelength case (Fig.~\ref{trajabsorb}c) is rather erratic due to many individual electron-ion collisions. 
The electron temperature $T_e$ towards the end of the laser pulse is $\approx 1.3$ (corresponding to $\approx 35$\,eV) while the quiver energy in the laser field $U_p$ is only $5.9\times 10^{-3}$.
Consequently, the motion of the electron inside the cluster is dominated by the thermal velocity rather than by the laser field. For the long wavelength instead, the free electron excursion is already greater than the (initial) cluster radius, the quiver energy $U_p$ is $0.38$, and the final electron temperature is of the same order of magnitude, namely  $T_e \approx 0.7$.

\subsubsection{Collision frequency}
\begin{figure}
\mbox{\qquad\qquad\qquad}\includegraphics[scale=0.48]{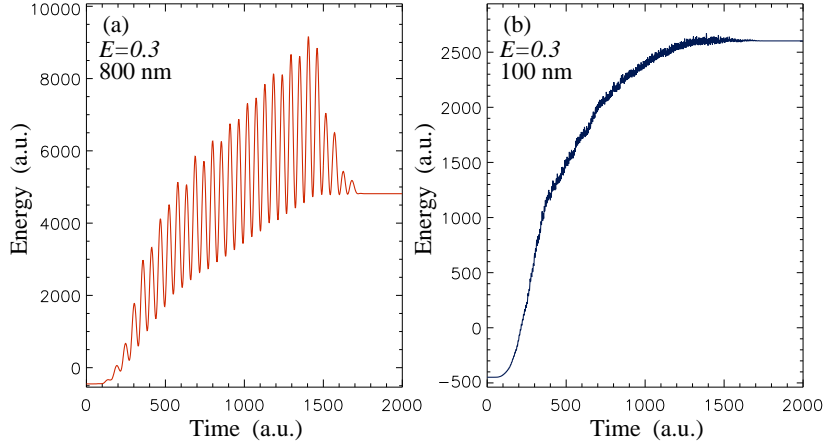}
\caption{\label{totenerg}  Total cluster energy vs time for $\Ehat=0.3$ at $800$ (a) and $100$\,nm (b). }
\end{figure}

For making comparison with analytical theories of inverse bremsstrahlung it is now attempted to calculate a time-averaged collision frequency $\langle\nu\rangle$ from the numerically determined total cluster energy $\energy(t)$. Absorption rate $\dot{\energy}$ and collision frequency are related through (see, e.g., \cite{mulser})
\beq \frac{\dot{\energy}}{N_e}=  \langle \dot{\vektr}\cdot\vektE\rangle = \langle \nu\dot{\vektr}^2\rangle \approx 2 \langle \nu \rangle U_p, \label{collfrequintro}\eeq
where $\langle\cdots\rangle$ indicates time-averaging over one laser cycle, $N_e$ is the number of electrons undergoing collisions, and $U_p=\Edach^2/(4\omega^2)$ is the quiver (or ponderomotive) energy. Calculating the collision frequency $\langle\nu\rangle$ is complicated by the fact that both $\dot{\energy}$ and $N_e$ vary in time. The total cluster energy as a function of time at $800$ and $100$\,nm is shown in Fig.~\ref{totenerg} for the case $\Ehat=0.3$~. It is obviously not possible to determine a constant absorption rate $\dot{\energy}$ from the overall slope of these curves which would be valid throughout the laser pulse. In particular at $100$\,nm the energy absorption strongly saturates because almost all active electrons leave the cluster. At low laser intensities, shortly after ionisation sets in at all, the behaviour is the opposite; then $\dot{\energy}$ increases during the pulse. 

Because of these difficulties related to the time-dependent electron density inside the cluster, an averaged single-electron absorption rate  was introduced as
\beq \dot{\epsilon}= \frac{\Delta\energy}{N_e T_\subsc{const}}, \eeq
where $T_\subsc{const}$ is the duration of the constant-intensity part of the trapezoidal laser pulse. For the number of electrons, $N_e=N_a \Zavcs(T)$ with $T$ the total pulse duration, was chosen at $800$\,nm and $N_e=N_a\Zavcs(T)-\Nouter(T)$ at $100$\,nm because free electrons may be driven back to the cluster in the former case while it is less likely in the latter. Taking the {\em final} values for $\Zavcs(t)$ and $\Nouter(t)$ may appear a too rough estimate. However, the average charge states, ion densities, and electron temperatures to be inserted in the analytical collision frequency formulas were also taken at time $T$ so that consistency is maintained.  

We compare the numerically determined collision frequency 
\beq \langle\nu\rangle=\frac{\dot{\epsilon}}{2U_p} \eeq
with 
\beq \langle\nu_{ei}^\subsc{(s,f)}\rangle = \frac{4\ 3^{3/2} (2\pi)^{1/2} n_i \Zavcs^2}{3 v^3} \ln\Lambda \label{silin} \eeq
where $n_i$ is the ion density, and $v$ is a typical electron velocity, $v=(3T_e+\Edach^2/\omega^2)^{1/2}$, which might be dominated by either the electron temperature $T_e$ or the driving laser field of amplitude $\Edach$. For the Coulomb logarithm either \cite{krainov}
\beq\ln\Lambda_s=\ln\left(\frac{4v^3v'}{\Zavcs^2 (v-v')^2 \ln^2C}\right) \quad \mbox{\rm (slow $e$)}\label{silinslow}\eeq
or
\beq \ln\Lambda_f=\ln\left|\frac{v+v'}{v-v'}\right| \quad \mbox{\rm (fast $e$)}\label{silinfast}\eeq
with $C$ the Euler constant and $v'=(v^2+2\omega)^{1/2}$ was taken.  The expression for fast electrons is expected to hold for $v\gg \Zavcs$ and $\omega\ll v^2$, while the one for slow electrons should be valid for $v\ll\Zavcs$, $\omega\ll v^3/\Zavcs\ll v^2$.

Since the condition $\omega\ll v^2$ is in general not fulfilled at $100$\,nm, we also make comparison with the expression derived in \cite{krainov}\footnote{Note that Eq.~(49) in \cite{krainov} differs from our Eq.~\reff{collfrequintro} by a factor of $2$. This translates to the expression for $\langle\nu_{ei}^\subsc{(K)}\rangle$.}
\beq \langle\nu_{ei}^\subsc{(K)}\rangle = \frac{8\pi^{3/2} n_i}{15\ 3^{5/6} (2T_e)^{1/2}} \left(\frac{2 \Zavcs^2}{\omega}\right)^{\!\!2/3} \frac{\Gamma(1/3)}{\Gamma(2/3)} \label{krainovcf} \eeq
which is supposed to be valid as long as $\omega\gg v^3/\Zavcs$. 

\begin{figure}
\mbox{\qquad\qquad\qquad}\includegraphics[scale=0.48]{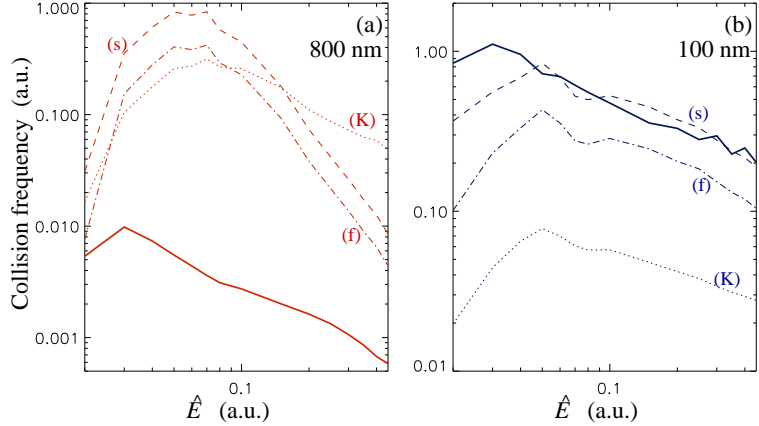}
\caption{\label{collcomp}  Collision frequencies vs peak field strength of the laser at $800$ (a) and $100$\,nm (b). Numerically determined collision frequency (solid, bold), $\langle\nu_{ei}^\subsc{(K)}\rangle$ Eq.~\reff{krainovcf} (dotted), $\langle\nu_{ei}^\subsc{(s)}\rangle$ Eqs.~(\ref{silin},\ref{silinslow}) (dashed), $\langle\nu_{ei}^\subsc{(f)}\rangle$ Eqs.~(\ref{silin},\ref{silinfast}) (dashed-dotted). }
\end{figure}

The result is shown in Fig.~\ref{collcomp}. It is seen that at $800$\,nm none of the electron-ion collision frequencies \reff{silin}--\reff{krainovcf} agrees with the numerical result, especially around $\Ehat\approx 0.06$ where both electron temperature and quiver energy are relatively low ($0.6$ and $0.28$, respectively). If the collision frequency $\langle\nu_{ei}\rangle$ was as high as predicted by these formulas, electron-ion collisions should clearly exceed the energy absorption by electron-cluster boundary collisions which, however, is not the case, as discussed in section~\ref{energabsorb}.  

The numerical result at $100$\,nm is in good agreement with $\langle\nu_{ei}^\subsc{(s)}\rangle$ for not too low laser intensities whereas  $\langle\nu_{ei}^\subsc{(K)}\rangle$ underestimates the numerically determined collision frequency. In Ref.~\cite{siedschlagIII} formula \reff{krainovcf} was found to overestimate collisional absorption which was attributed to screening effects not included in the derivation of \reff{krainovcf}. In fact, the Debye-length $\lambda_D=[T_e/(4\pi n_e]^{1/2}$ is of the same size or even smaller than the interionic distance $n_i^{-1/3}$ so that one would expect \reff{krainovcf} to overestimate the collision frequency---which it does not in our case. 

The reason for the discrepancies between the simulation results for the collision frequency and the analytical values is not clear (at least to the author). Only if one allows for an unreasonably generous adjustment of the effective charge states (commonly justified with screening of the nuclear charge) acceptable agreement could be achieved in the 800\,nm case.  Also unclear is why, at 100\,nm, eqs.~(\ref{silin},\ref{silinslow}) work better than expected while the seemingly more appropriate formula \reff{krainovcf} gives poor agreement. 

In Ref.~\cite{santra} enhanced laser energy absorption was attributed to electrons scattering on an effective potential that changes from Coulombic with the full nuclear charge in the interior to a Debye-screened potential with the actual charge state in the exterior. Electrons with small impact parameter thus ``see'' a nuclear charge higher than the actual charge state. This effect, of course, is not specific to laser-cluster interaction but is a general issue in plasma physics. It is automatically included---may it be important or not---in our numerical calculations (at least as long as the number of actively treated electrons is sufficient) while it is not incorporated in molecular dynamics simulations dealing with rates for inner ionisation and effective charge states only.

In our work, ionisation ignition was found to be the dominant ionisation mechanism in clusters also at short wavelengths. This agrees with results presented in \cite{siedschlagIII}. Ionisation ignition, upon relying on outer ionisation, is a finite-size effect not considered in \cite{santra}.  It is the mechanism for outer ionisation that depends on the laser wavelength and ranges from collective electron dynamics (long wavelength) to collisional absorption and thermionic emission (short wavelength). Hence, inverse bremsstrahlung only {\em indirectly}, via outer ionisation, leads to the generation of high charge states in clusters.

\section{Conclusions} \label{concl}
Results from numerical simulations of a semi-classical Xe$_{54}$ model cluster in short and intense laser pulses at $800$ and $100$\,nm wavelength were presented and compared with the corresponding single atom results. The inner-atomic dynamics was treated explicitly so that no rates for field ionisation, collisional ionisation, and recombination had to be employed but all physical effects, although classical ones only, were included exactly. 

It was found that (dynamical) ionisation ignition is the by far dominating mechanism behind enhanced ionisation of atoms inside clusters at both long and short wavelength. The outer ionisation mechanism depends on the laser wavelength.
At long wavelengths, when the electron excursion is comparable to the cluster radius, laser energy absorption proceeds mainly through electron-cluster (boundary) collisions while at short wavelengths, where the electron dynamics is dominated by the thermal rather than the quiver velocity, inverse bremsstrahlung prevails.

Collisional ionisation is insignificant. However, under the conditions studied in this paper the laser field turns the cluster into a strongly coupled plasma  where inelastic processes such as electron capture and reemission occur with high probability. As a consequence, the charge states remained moderate although all active electrons in the simulation left their parent atom.

The charge state distributions of the cluster ions are broader, ranging from the highest charge state generally produced at the cluster boundary down to the lowest one produced in the cluster centre. At short wavelengths the small electron quiver amplitude and the moderate temperatures reached in short pulses hamper outer ionisation. As a result, after cluster expansion the final, mean charge state may lag behind the corresponding single atom result.

Recent work employing the ``particle-in-cell'' simulation technique \cite{jungreuth} indicates that at long wavelengths dynamical ionisation ignition (dubbed ``polarisation enhanced ionisation'' there) and electron-cluster boundary collisions (called ``laser dephasing heating'' in \cite{jungreuth}) remain to be the dominant ionisation and absorption mechanisms in big clusters of several $10^5$ atoms as well. 

Finally, the question arises whether the explicit treatment of the inner-atomic dynamics leads to results different from those obtained with simpler approaches. On one hand, the method used in this work, although semi-classical only, is fully self-consistent and clearly superior to approaches where field ionisation, recombination, collisional ionisation, etc.\ are treated through rate equations. Accurate rates for these processes in non-equilibrium laser plasmas are hardly available, and the different processes cannot even be properly separated. On the other hand, there is a type of simulations where the outermost bound electron is considered explicitly while the deeper bound ones are accounted for by an effective nuclear charge of the parent atom. The parent atom charge state is increased when the electron leaves, and the same procedure starts for the next bound electron. These kind of simulations which enforce ``purely sequential parent atom ionisation'' (PSPAI) should lead to similar results as compared to ours since non-sequential ionisation does not play a major role. In particular, there should be the same important difference between the physically meaningful charge state \reff{cs} and the non-observable parent atom ionisation.

If very high charge states are created the method used in the present work becomes more rapidly inefficient than those neglecting the multi-electron, inner-atomic dynamics because too many electrons per atom need to be simulated {\em from the beginning}.  In such cases the ``PSPAI approach'' described in the previous paragraph is the method of choice, possibly amended to account for explicit recombination so that the effective nuclear charges and the number of electrons in the simulation will eventually {\em decrease} while the cluster expands.

\ack
This work was supported by the  Deutsche Forschungsgemeinschaft. Fruitful discussions with Andrea Macchi are gratefully acknowledged. The author thanks Andreas Becker and Hubertus Wabnitz for providing the experimental data of \cite{laro} and \cite{wabnitzXeatoms} in electronic form.  The permission to run our codes  on the Linux cluster at PC$^2$ in Paderborn, Germany, is highly appreciated.

\end{document}